\documentclass[twocolumn,10pt]{revtex4}
\usepackage{graphicx}
\usepackage{amssymb,amsmath,amsfonts,amsthm}

\newcommand{\dR}{\mathbb R}

\newcommand{\dZ}{\mathbb Z}

\begin{document}

\title{Quantum of volume in de Sitter space}

\author{Jakub Mielczarek}
\email{jakub.mielczarek@uj.edu.pl}
\affiliation{Astronomical Observatory, Jagiellonian University, 30-244 Krak\'ow, Orla 171, Poland}

\author{W{\l}odzimierz Piechocki}
\email{piech@fuw.edu.pl}
\affiliation{Theoretical Physics Department, Institute for Nuclear
Studies, Ho{\.z}a 69, 00-681 Warszawa, Poland }

\begin{abstract}
We apply the  nonstandard loop quantum cosmology method
to quantize a flat Friedmann-Robertson-Walker cosmological 
model with a free scalar field and the cosmological constant 
$\Lambda>0$. Modification of the Hamiltonian in terms of loop 
geometry parametrized by a length $\lambda$ introduces a scale 
dependence of the model. The spectrum of the volume operator 
is discrete and depends on $\Lambda$. Relating quantum of 
the volume with an elementary lattice cell leads to an explicit 
dependence of $\Lambda$ on $\lambda$. Based on this assumption, 
we investigate the possibility of interpreting $\Lambda$ as a 
running constant.
\end{abstract}

\maketitle

\section{Introduction}

The cosmological observations, interpreted  in terms of the
Friedmann-Robertson-Walker (FRW) model, suggest that the Universe
underwent an accelerated expansion at least twice in its history.
The first one is called the cosmic inflation and it is specific to
the very early Universe \cite{Guth:1980zm}. The second one is the
presently observed dark energy domination area
\cite{Copeland:2006wr}.  In both cases the evolution can be
explained, to some extent, by a constant or nearly-constant
contribution to the energy density modeled by the so-called
cosmological constant $\Lambda$. In gravitational physics one
usually treats $\Lambda$ as a constant, like Newton's constant
$G$, that cannot be derived from first principles
\cite{Bianchi:2010uw}. In cosmology one usually calls $\Lambda$,
in the case of FRW model, the simplest  model of dark energy
\cite{Copeland:2006wr} and one tries to find its nature.

In this paper, we study a possible link between $\Lambda$ and the
{\it discrete} structure of space associated with quantum theories
of gravity. Loop quantum cosmology (LQC) is a suitable
framework to address this issue.  In particular, we quantize the
space {\it volume} function to find the relation between $\Lambda$
and the length $\lambda$ specifying loop geometry underlying LQC.
In what follows, we consider a flat FRW model with the
cosmological constant and a free scalar field.

In standard LQC one applies the Dirac quantization method
\cite{Ashtekar:2003hd,Bojowald:2006da}, which postpones solution
of the Hamiltonian constraints to the quantum level. In the
nonstandard LQC  one solves constraints already at the classical
level and quantization is carried out on the {\it physical} phase
space. It is called the reduced phase space quantization method.
This approach has been successfully applied to the FRW
\cite{Dzierzak:2009ip,Malkiewicz:2009qv} and the Bianchi I
\cite{Dzierzak:2009dj,Malkiewicz:2010py} models with a free scalar
field. Recently, the constraints for the FRW model with a free
scalar field and the cosmological constant have been solved too
\cite{Mielczarek:2010rq}. The physical observables like the volume
and the energy density of matter field have been analyzed, for
both positive and negative $\Lambda$, at the classical level. In
what follows we  present quantization of the volume observable for
the case of $\Lambda > 0$.

\section{Hamiltonian}

The Hamiltonian of the model {\it modified} by the holonomy around
a loop  takes the form \cite{Mielczarek:2010rq,Bentivegna:2008bg}
\begin{eqnarray}
H^{(\lambda)} &=&  -\frac{vN}{32\pi^2 G^2\gamma^3\lambda^3} \sum_{ijk}
\epsilon^{ijk} \text{tr} \left[h_{\Box_{ij}}h_k \left\{ (h_k)^{-1}
,v\right\}\right]   \nonumber \\
&+&N\frac{p^2_{\phi}}{2 v} \label{hgl}+N\frac{v\Lambda}{8\pi G}
\approx 0, \label{Hamiltonian}
\end{eqnarray}
where "$\approx$" reminds that the Hamiltonian is a {\it constraint}
of the system, $\gamma$ is the Barbero-Immirzi parameter, $G$ is
Newton's constant, $p_\phi$ is the conjugate momentum of the
scalar field $\phi$, $N$ is the lapse function, $h_{\Box_{ij}} =
h_i h_j (h_i)^{-1} (h_j)^{-1}$ is the holonomy around the square
loop $\Box_{ij}$,  $h_i = \cos\left(\lambda
\beta/2\right)\mathbb{I}+2\sin\left( \lambda \beta/2\right)
\tau_i$ is the holonomy in the i-th direction, and where
$\tau_i=-\frac{i}{2} \sigma_i$ ($\sigma_i$ are the Pauli
matrices). The variable $v$ is the volume of a piece of space
$\mathcal{V}\subset \dR^3$ (we assume that the spacelike part of
spacetime has $\dR^3$ topology) defined  as follows: $v =
\int_\mathcal{V} dx_1 dx_2 dx_3 \sqrt {det \,q_{ab}} $, where
$\,q_{ab} dx^a dx^b:= a^2 (dx_1^2 + dx_2^2 + dx_3^2)\,$ defines
the FRW metric ($a$ is the  scale factor). The variable $\beta$ is
related with the Hubble parameter. The variable $\lambda$, having
the dimension of a length, is a {\it free} parameter of the theory
\cite{Dzierzak:2008dy}.  The Hamiltonian (\ref{Hamiltonian}) may
be seen as the {\it lattice} discretized version of the classical
expression, where  $\lambda$ plays the role of a lattice constant
and $\lambda^3$ is the volume of an elementary cubic cell. While
$\lambda \rightarrow 0$, the nonmodified general relativity
Hamiltonian is recovered.

In LQC the gravitational degrees of freedom are represented by
holonomies and fluxes. A holonomy used in LQC is a connection
integrated  along spatial elementary loop.  Making use of
holonomies enables, roughly speaking, the resolution of the
cosmological singularity problem \cite{Malkiewicz:2009qv}. Thus,
they are of primary importance. In the flat FRW model it is enough
to use one holonomy parametrized by $\lambda$. Equation
(\ref{hgl}) {\it does not} define, in the nonstandard LQC
\cite{Dzierzak:2009ip,Malkiewicz:2009qv}, an effective
semiclassical Hamiltonian, but a classical {\it modified}
Hamiltonian. The modification is specified by the value of
$\lambda$.  The bigger $\lambda$, the bigger the smearing of the
holonomy variable in the Hamiltonian (\ref{hgl}). Because of the form
of (\ref{hgl}),  we propose to interpret $\Lambda$ to be a {\it
coupling} constant depending on the {\it smearing} $\lambda$. We
promote this interpretation in Sec. VI to relate $\Lambda$
with $\lambda$.

The {\it kinematical} phase space is parametrized by four
canonical variables $(v,\beta,p_{\phi},\phi)$.  The imposition of
constraint (\ref{hgl}) leads to the {\it physical} phase space
parametrized  by two elementary observables $\mathcal{O}_1$ and
$\mathcal{O}_2$, satisfying the algebra $\left\{
\mathcal{O}_2,\mathcal{O}_1 \right\} = 1$.    The dynamics of the
model is traced by the scalar field $\phi$ which plays the role of
an intrinsic time (see, \cite{Mielczarek:2010rq} for more
details).

\section{Observables}

It is shown in
\cite{Mielczarek:2010rq} that the elementary observables of our
model are
\begin{eqnarray}
\mathcal{O}_1 &=& p_{\phi}, \\
\mathcal{O}_2 &=& \phi+\frac{\text{sgn}(p_{\phi})}{\sqrt{12\pi G
\delta}} \frac{1}{i} \left[ F\left( \beta\lambda \left|
\frac{1}{\delta} \right. \right)-F\left(\frac{\pi}{2}\left|\frac{1}
{\delta}\right.\right)\right]. \label{O2dS}
\end{eqnarray}
Here $F(\cdot | \cdot)$  is the Jacobi elliptic function of the
first kind, and $\delta:=\frac{1}{3} \Lambda \gamma^2 \lambda^2$.
For this model the allowed values of the parameter $\delta$ are in
the set $[0,1]$. The lower limit correspond to $\Lambda=0$ case
while the upper limit is the dynamically allowed value
corresponding to $\Lambda = \frac{3}{\gamma^2\lambda^2} =:
\Lambda_{\text{c}}$ (see, \cite{Mielczarek:2008zv} for more
details).

\begin{figure}[ht!]
\centering
\includegraphics[width=7cm,angle=0]{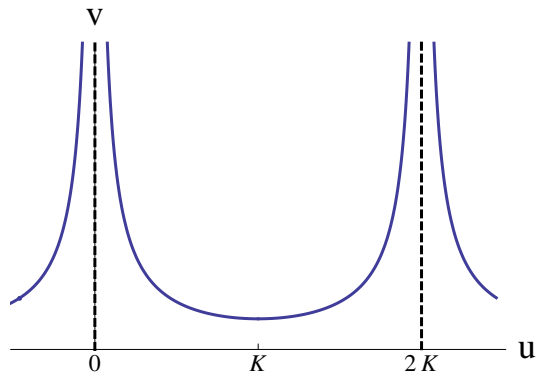}
\caption{Evolution of volume $v$ as a function of the parameter
$u$. The function is periodic in variable $u$ with the period
equal to $2K$. The solutions in different periods are separated by
vertical asymptotes.} \label{v}
\end{figure}

An expression for the volume function reads
\cite{Mielczarek:2010rq}
\begin{equation}
v = \frac{|\mathcal{O}_1|}{\sqrt{2\rho_c(1-\delta)}} \frac{1}
{\left|\text{sn}\left(u \left| 1-\frac{1}{\delta} \right.
\right)\right|} =: |w|, \label{vu}
\end{equation}
where
\begin{equation}
u := \sqrt{12\pi
G\delta}(\mathcal{O}_2-\phi)+K,~~~~\rho_{\text{c}} :=
\frac{3}{8\pi G \gamma^2 \lambda^2},
\end{equation}
and where $K:=F(\pi/2 \left|1-1/\delta\right.)$. Function
$\text{sn}(\cdot | \cdot)$ is the elliptic sinus function defined
as $\text{sn}(u|m):=\sin \text{am} (u|m)$, where $\text{am}
(u|m):=F^{-1}(u|m)$ is the amplitude of the $F$ elliptic function.
We have also used here a definition of the critical energy
density. In Fig. \ref{v} we plot an evolution of the volume given
by Eq. (\ref{vu}).

\section{Quantization}

Quantization of the system follows the method presented in
\cite{Malkiewicz:2009qv}.  We choose the
Schr\"{o}dinger representation for the classical algebra $\left\{
\mathcal{O}_2,\mathcal{O}_1 \right\} = 1$ in the form
\begin{equation}
\hat{\mathcal{O}}_1\psi(x):= -i\hslash \frac{d}{dx}\psi(x),~~~~
\hat{\mathcal{O}}_2\psi(x) :=  x\psi(x),
\end{equation}
where $\psi \in L^2 (\mathbb{R})$, so we get
$\left[\hat{\mathcal{O}}_2,\hat{\mathcal{O}}_1 \right] = i\hslash
\,\mathbb{I}$. Quantization of $w$ may be done in a standard way
as follows
\begin{equation}
\hat{w}  := \frac{1}{\sqrt{2\rho_c(1-\delta)}}
\frac{1}{2}\Big(\hat{\mathcal{O}}_1 \;g(\hat{\mathcal{O}}_2-\phi)
+ g(\hat{\mathcal{O}}_2-\phi)\;\hat{\mathcal{O}}_1\Big),
\end{equation}
where
\begin{equation}
g(\hat{\mathcal{O}}_2-\phi):=\frac{1}{\text{sn}\left( \sqrt{12\pi G\delta}
(\hat{\mathcal{O}}_2-\phi)+K\left|1-\frac{1}{\delta} \right. \right)}.
\end{equation}

We are looking for the solution to the eigenvalue problem of the
operator $\hat{w}$
\begin{equation}\label{eigen1}
\hat{w} \;\psi_b(x) = b \; \psi_b(x),~~~b \in \mathbb{R} .
\end{equation}
This leads to the equation for eigenfunctions in the form
\begin{eqnarray}
\frac{d}{dx} \left( g(x-\phi) \psi_b(x)\right)
&+& g(x-\phi)\frac{d\psi_b(x)}{dx} \nonumber \\
&=& b \frac{i}{\hslash} 2 \sqrt{2\rho_c(1-\delta)} \psi_b(x).
\nonumber
\end{eqnarray}
A general  solution to this equation reads
\begin{equation}
\psi_b(u)=\psi_0 \sqrt{\text{sn} \left(u\left|1-\frac{1}{\delta}
\right.\right)} \exp \left\{ \frac{i}{\hslash}
\frac{b\sqrt{\rho_{\text{c}}}}{\sqrt{6\pi G}} \Theta(u)\right\},
\end{equation}
where
\begin{equation}
\Theta(u) := \arctan\left\{ - \sqrt{\frac{1-\delta}{\delta}}
\frac{\text{cn}\left(u \left|1-\frac{1}{\delta} \right.\right)}
{\text{dn}\left(u \left|1-\frac{1}{\delta} \right.\right)}
\right\},
\end{equation}
and where the elliptic functions are defined as follows:
$\text{cn}(u|m)= \cos\text{am} (u|m)$, and
$\text{dn}(u|m)=\sqrt{1-m\ \text{sn}^2(u|m)}$. The normalization
condition for the eigenfunctions reads
\begin{eqnarray}
1 =\langle \psi_b | \psi_b \rangle  =
\frac{|\psi_0|^2}{\sqrt{12\pi G\delta}} \int_{0}^{2K}du \
\text{sn} \left(u\left|1-\frac{1}{\delta} \right.\right).
\end{eqnarray}
We integrate over one period of evolution as other periods
correspond to the same model of the universe
\cite{Mielczarek:2010rq}.  Thus, the normalization factor is found
to be
\begin{equation}
\psi_0 = \sqrt{\frac{\sqrt{3\pi G(1-\delta)}}{\arctan\left(\sqrt{\frac{1}
{\delta}-1} \right)}}.
\end{equation}
To find an orthonormal set of eigenfunctions, we calculate
\begin{eqnarray}
\langle \psi_b | \psi_a \rangle &=&\frac{1}{\sqrt{12\pi G\delta}}\int_{0}^{2K}
du \  \psi^*_b(u) \psi_a(u) \nonumber \\
&=&\frac{|\psi_0|^2}{\sqrt{3\pi G\delta}} \frac{\sin \left[ C
\arctan\sqrt {\frac{1}{\delta}-1} \right]}{C
\sqrt{\frac{1}{\delta}-1}},
\end{eqnarray}
where
\begin{equation}
C :=\frac{1}{\hslash}
\frac{(b-a)\sqrt{\rho_{\text{c}}}}{\sqrt{6\pi G}}.
\end{equation}

\section{Quanta of volume}

The orthogonality condition
$\langle \psi_b | \psi_a \rangle=0$ leads to
\begin{equation}\label{eval}
b = a + m\;\Delta_{\delta},~~~a \in \mathbb{R},~~~m \in
\mathbb{Z},
\end{equation}
where
\begin{equation}\label{gap}
\Delta_{\delta} :=8\pi G \gamma \lambda \hslash  \frac{\pi/2}
{\arctan\sqrt{\frac{1}{\delta}-1}}.
\end{equation}
Therefore, the eigenvalues of $\hat{v}$, due to (\ref{vu}), are:
$c = |a + m\;\Delta_{\delta}|$. Thus, the spectrum is {\it
discrete}, which is of a basic importance for the rest our paper.

The space $\mathcal{F}_a:=\{~\psi_b\;|\; b = a +
m\;\Delta_{\delta};~m\in \dZ;~b \in \dR~\}$ is {\it orthonormal}.
Each subspace $\mathcal{F}_a \subset L^2(\dR)$ spans a pre-Hilbert
space.  The completion of $D_a(\hat{w}):= span
\;\mathcal{F}_a,~\forall a \in \dR$, leads to $L^2(\dR)$. One may
prove (in analogy to the corresponding proof in
\cite{Malkiewicz:2009qv}) that the operator $\hat{w}$ is
essentially {\it self-adjoint} on $D_a(\hat{w}),~\forall a \in
\dR$.

The expression (\ref{gap}) specifies the minimum gap in the
spectrum of the volume operator (for $a = 0$) in terms of the
cosmological constant $\Lambda$ so it defines a {\it quantum} of
the volume. It is interesting to examine the limit when $\delta
\rightarrow 0 $ ($\Lambda \rightarrow 0$). Because of the relation
$\lim_{\delta\rightarrow 0^+}\arctan\sqrt{\frac{1}{\delta}-1}= \frac{\pi}{2}$,
we get $\lim_{\delta\rightarrow 0^+}\Delta_{\delta} =8\pi G \gamma
\lambda \hslash  =: \Delta. $ This is precisely an expression that
has been found earlier \cite{Malkiewicz:2009qv} in the case
$\Lambda=0$, which proves the consistency of our results. While
approaching $\delta\rightarrow 1$ ($\Lambda\rightarrow
\Lambda_{\text{c}}$) we find that $\Delta_{\delta} \rightarrow
\infty$.

\begin{figure}[ht!]
\centering
\includegraphics[width=7cm,angle=0]{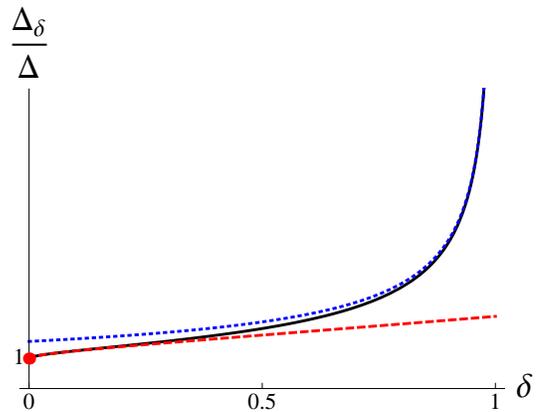}
\caption{Ratio $\Delta_{\delta}/\Delta$ as a function of  $\delta$ (black line).
We see that $\lim_{\delta\rightarrow 0}\Delta_{\delta}/\Delta =1$
as well as $\lim_{\delta\rightarrow 1}\Delta_{\delta}/\Delta =\infty$.
The (red) dot represents the case with $\Lambda=0$.}
\label{vdS}
\end{figure}

In Fig. \ref{vdS}  we plot the ratio  $\Delta_{\delta}/\Delta$
as a function of  $\delta$.
One may easily verify that for $\delta\rightarrow 0$ we have
\begin{equation}
\frac{\Delta_{\delta}}{\Delta} = 1+\frac{2\sqrt{\delta}}{\pi}+
\frac{4\delta}{\pi^2}+\mathcal{O}(\delta^{3/2}) \label{approx1},
\end{equation}
whereas  for $\delta\rightarrow 1$ we get
\begin{equation}
\frac{\Delta_{\delta}}{\Delta} = \frac{\pi/2}{\sqrt{1-\delta}}+
\mathcal{O}(\sqrt{1-\delta}). \label{approx2}
\end{equation}
These approximations are also plotted in Fig. \ref{vdS}.  The
dashed (red) line is an approximation (\ref{approx1}) while dotted
(blue) line is an approximation (\ref{approx2}).

It is instructive to check the value of $\Delta_{\delta}$ 
in the observed Universe. The cosmological constant $\Lambda$ can
be related with the observed dark energy, which dominates the
energy density of the Universe. In this case one can rewrite the
definition of $\delta$ parameter in the form
\begin{equation}
\delta = \Omega_{\Lambda} H^2_0 \gamma^2 \lambda^2,
\end{equation}
where $\Omega_{\Lambda}$ is the fractional density of the cosmological constant,
and $H_0$ is the present value of the Hubble parameter.
The five years of observations of the WMAP satellite yield
$\Omega_{\Lambda}=0.742 \pm 0.030$ and $H_0=71.9^{+2.6}_{-2.7} \
\text{km} \ \text{s}^{-1}\ \text{Mpc}^{-1}$ \cite{Dunkley:2008ie}.
Assuming that $\lambda=l_{\text{Pl}}$, where $\l_{\text{Pl}}$ is
the Planck length and $\gamma=0.2375\,$ \cite{Meissner:2004ju}, we find
\begin{equation}
\delta = 6.6 \cdot 10^{-124}.
\end{equation}
Because this value is extremely small, the value of
$\Delta_{\delta}$ with the high precision overlaps with $\Delta$,
obtained in the $\Lambda=0$ limit. Therefore $\Delta_{\delta}  =
8\pi \gamma \l_{\text{Pl}}^3$, where we have assumed
$\lambda=l_{\text{Pl}}$, as previously. With $\gamma=0.2375$, the
quanta of volume takes a value $\Delta_{\delta}  \approx  6 \
v_{\text{Pl}}$, where $v_{\text{Pl}}:=l_{\text{Pl}}^3$ is a Planck
volume.

\section{Running of $\Lambda$?}

We propose to relate the quantum of the volume $\Delta_{\delta}$, defined by (\ref{gap}),
with the volume $\lambda^3$  of an {\it elementary} lattice cell
as follows
\begin{equation}\label{post}
\Delta_{\delta} =    \lambda^3 ,
\end{equation}
which leads to the equation
\begin{equation}
\lambda^2=8\pi\gamma l^2_{\text{Pl}}\,
\frac{\pi/2}{\arctan\sqrt{\frac{1}{\delta}-1}}. \label{l2}
\end{equation}

Because $0 \leq \arctan\sqrt{\frac{1}{\delta}-1} \leq \pi/2$, it
is clear that $\lambda \in [\lambda_0,\infty[$, where $\lambda_0
:=\sqrt{8\pi \gamma }\ l_{\text{Pl}}$.  Therefore  (\ref{l2})
leads to the constraint on the value of $\lambda$ from below. The
minimum value $\lambda_0$ corresponds to the case $\Lambda=0$.

Equation  (\ref{post}) is analogous to the equation postulated in
standard LQC for the determination of the {\it minimum} length of
a loop along which the holonomy is defined. One requires that the
area of the minimum loop in LQC equals the smallest nonzero
eigenvalue of the area operator of loop quantum gravity (LQG)
(see, e.g. \cite{Ashtekar:2003hd,Ashtekar:2006wn}). However, in
our case we make the postulate at the {\it physical} sector of
LQC, contrary to the case of \cite{Ashtekar:2006wn} where one
compares the corresponding quantities from LQC and LQG in the {\it
kinematical} sector of both theories.

Equation (\ref{l2})  can be inverted into the form
\begin{equation}\label{lam}
\Lambda=\Lambda_{\text{c}} \cos^2\left(\frac{\pi}{2}
\frac{\lambda^2_0}{\lambda^2} \right),
\end{equation}
where $\lambda \in [\lambda_0,\infty[$ as shown previously. This
way we have turned the cosmological constant $\Lambda$ into a {\it
variable} cosmological constant, a function depending explicitly
on $\lambda$.

\begin{figure}[ht!]
\centering
\includegraphics[width=7cm,angle=0]{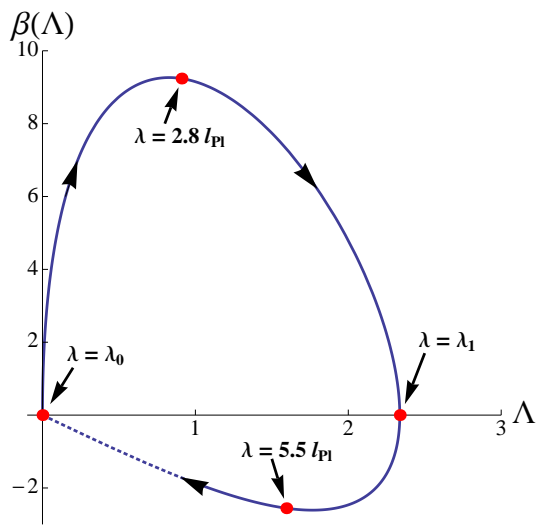}
\caption{Scale dependence of  $\Lambda$. The arrows
indicate flow from UV ($\lambda=\lambda_0$) to IR limit
($\lambda\rightarrow \infty$). The dotted part of the curve
denotes the region of large $\lambda$ ($\lambda\gg\lambda_1$)
where considerations become speculative.  In the plot $\lambda_0 =
2.4\,l_{\text{Pl}}$ and  $\lambda_1 = 3.8\,l_{\text{Pl}}.$}
\label{RGF}
\end{figure}

One may  treat the relation  (\ref{lam}) as an analogue of the
expression for the vacuum energy obtained within the framework of
quantum field theory. Namely, the vacuum energy may be estimated
by summing up energies of the zero modes down to some cut-off
length scale $\tilde{\lambda}$. Therefore, for the cosmological
constant interpreted as a vacuum energy, one gets
\begin{eqnarray}
\Lambda &=& 8\pi l^2_{\text{Pl}} \rho_{\text{vac}} \sim
 l^2_{\text{Pl}}   (n_{\text{b}}-2n_{\text{f}}) \int_0^{1/\tilde{\lambda}}k^3   dk  \nonumber \\
 &=& (n_{\text{b}}-2n_{\text{f}})
\frac{l^2_{\text{Pl}}}{4\tilde{\lambda}^4}.
\end{eqnarray}
Here, $n_{\text{b}}$ and $n_{\text{f}}$ is the number of bosons
and fermions, respectively. The value of cosmological constant is
explicitly expressed in terms of some minimal length scale
$\tilde{\lambda}$. Substitution $\tilde{\lambda} \sim
l_{\text{Pl}}$ leads however to the known problem of the large
vale of cosmological constant, $\Lambda \sim m^2_{\text{Pl}}$
(that can be "solved" by imposing the condition of supersymmetry
$n_{\text{b}}=2n_{\text{f}}$, which leads to
$\rho_{\text{vac}}=0$).  Now, let us make use of the relation
(\ref{lam}). It is helpful to expand  (\ref{lam}) in terms of
$\epsilon := \frac{\lambda-\lambda_0}{\lambda_0}$, where
$\lambda_0 = \sqrt{8\pi \gamma} l_{\text{Pl}}$. So we get
\begin{equation}
\Lambda = \frac{3\pi}{8 \gamma^3} m^2_{\text{Pl}} \epsilon^2
\left(1-5\epsilon+\mathcal{O}(\epsilon^2)\right).
\end{equation}
We can see that the value of our cosmological constant is also
proportional to $m^2_{\text{Pl}}$. However, there is additional
factor $\epsilon^2$, that can be used to match $\Lambda$  with the
observed value. From the observations we have that
$\frac{\Lambda}{m^2_{\text{Pl}}}\approx 10^{-123}$, which leads to
$\epsilon \approx 10^{-63}$. The value of $\lambda$ must be
therefore very close to $\lambda_0 $ in order to reproduce
observed value of the cosmological constant. Therefore, in
contrast to the field theoretical approach to $\Lambda$, in our
case, the observed value of $\Lambda$ can be obtained for
discretization scale $\lambda \approx l_{\text{Pl}}$! To examine
the relation $\Lambda = \Lambda (\lambda)$ in more details, we
introduce the $\beta-$function (it should be not confused with the
canonical variable $\beta$) as follows
\begin{equation}
\beta(\Lambda):= \lambda \frac{d\Lambda}{d\lambda}.
\end{equation}
This function is defined in analogy to the $\beta-$function in the
renormalization group theory \cite{KGW}. However, in our case,
this function is only a tool that we use to visualize the scale
dependence. We plot the resulting $\beta-$function in Fig.
\ref{RGF}.

One can see that on the very small scale,  $\Lambda$ grows from
$0$ at the first fixed point $\lambda_0$, and approaches the
maximum $\Lambda \approx 2.3\, m^2_{\text{Pl}}$ at the second
fixed point $\lambda_1$.  In the high energy domain (small
$\lambda$) the value of $\Lambda$ grows from zero to the Planck
scale level. While approaching the largest scales with growing
$\lambda$, the value of the cosmological constant tends to zero
again, $\lim_{\lambda \rightarrow \infty} \Lambda(\lambda) = 0$.
However, we should keep in mind that as $\lambda$ grows, our
lattice interpretation  becomes less and less justified, and it
does not make sense as $\lambda \rightarrow \infty$.

Using the above arguing one may suggest that, in a sense,
cosmological constant can be interpreted as a running coupling
constant. In such an interpretation, $\lambda$ is treated as a
scale at which the high energy degrees of freedom are integrated.
However, systematic verification of this suggestion is needed. We
believe that our considerations give some hints which can
stimulate further investigations in this direction.

\section{Conclusions}

In this paper we present quantization of the FRW model with the
cosmological constant and the free scalar field in the framework
of the nonstandard loop quantum cosmology.  We investigate
spectral properties of the volume operator.  We find the
eigenfunctions as well as corresponding spectrum which is shown to
be discrete. Based on this, we find an expression  for the
elementary quanta of volume as a function of cosmological constant
$\Lambda$. In the limit $\Lambda \rightarrow 0$,  our formula
reduces to the expression that was found earlier in
\cite{Malkiewicz:2009qv} for the $\Lambda=0$ case.

Postulating the relation of the quantum of a space volume with the
volume of a lattice cell, we express the cosmological constant
$\Lambda$ in terms of $\lambda$. Based on this we {\it suggest}
that $\Lambda$ may play  the role of a running constant similar to
the coupling constant in QCD which decreases with increasing
energy scale. In the IR limit (where our model is little
justified), $\Lambda$ behaves similarly to the fine structure
constant in QED. We are conscious that our dependence of $\Lambda$
on $\lambda$ is not a renormalization group flow in any {\it
standard} sense. The renormalization group in quantum field theory
comes from integrating out high energy degrees of freedom and
requiring that this be compensated  by changes in the coupling
constants.  We simply put forward a {\it hypothesis} of
interpreting $\Lambda$ as a running constant to be verified by
future studies of the nature  of the cosmological constant.

The free parameter $\lambda$, of the classical level, can be
restricted after quantization by making use of (\ref{l2}). As it
was shown earlier, the {\it lowest} allowed value of $\lambda$
equals $\lambda_0$. For this value,  the corresponding critical
energy density $\rho_{\text{c}}(\lambda_0) = \frac{3
m^4_{\text{Pl}}}{64\pi^2 \gamma^3} \approx  0.35 \
m^4_{\text{Pl}}$, for $\gamma = 0.2375$ (see,
\cite{Meissner:2004ju}). This is a bit lower than the value
obtained within the standard LQC,  $\rho_{\text{c}} \approx 0.82 \
m^4_{\text{Pl}}$ \cite{Ashtekar:2006wn}.  The value of
$\Lambda_{\text{c}}$ at $\lambda=\lambda_0$ is given by
$\Lambda_{\text{c}}(\lambda_0)= \frac{3 m^2_{\text{Pl}}}{8\pi
\gamma^3} \approx  8.9  \ m^2_{\text{Pl}}$, that is comparable
with the value obtained in the standard LQC,
$\Lambda_{\text{c}}\approx 10.3  \ m^2_{\text{Pl}}$
\cite{Kaminski:2009pc}.

It is clearly seen that we could carry out the analyses due to an
application of LQC method in the reduced phase space quantization
version. In this approach an elementary length $\lambda$ is a free
parameter. Such an interpretation cannot be done within LQC which
is based on Dirac's quantization with the parameter $\lambda$
having a fixed value \cite{Ashtekar:2006wn}.

In the standard LQC an important issue is choosing the correct
{\it lattice} refinement that appears in the context of the
implementation of the Hamiltonian constraint at the quantum level
defining an evolution of a quantum system. Ignoring this problem
leads to serious {\it instabilities} in the continuum
semiclassical limit. Addressing properly this issue has required
much effort (see, \cite{MS, Corichi:2008zb} and references
therein). The nonstandard LQC is {\it free} of this problem as the
constraint is being solved already at the {\it classical} level
leading to the physical phase space. As the result one imposes
quantum rules into the system without constraints, but with
sophisticated phase space. An evolution of a quantum system, in
our method,  is defined in terms of a self-adjoint so-called true
Hamiltonian via Stone's theorem \cite{Malkiewicz:2010py, JW}.

The refinement is a choice of the scale factor dependence of the
link length for the lattice states.  In particular, the scaling
parameter  is considered in the form $\tilde{\mu} \propto v^{n}$,
where $n \in [-1/3,0]$. The case $n=0$ corresponds to the
so-called {\it old} quantization scheme ($\mu_0$-scheme) which
leads to a wrong semiclassical behavior. It was shown in Ref.
\cite{Mielczarek:2008zv} that for some positive values of
cosmological $\Lambda$ one obtain the oscillatory behavior, which
is not expected to occur at classical level. The case $n=-1/3$
corresponds to the so-called {\it new} quantization scheme
($\bar{\mu}$-scheme). In this case an area of the loop,
$\text{Ar}_{\Box_{ij}}$, remains constant during an evolution of
the system. It is so because $a \propto v^{1/3}$ and  $\tilde{\mu}
\propto a^{-1} \propto v^{-1/3}$, thus  $\text{Ar}_{\Box_{ij}}
\propto a^2 \tilde{\mu}^2 = \Delta =$ const, where $\Delta $ is
the area gap derived within LQG. In this particular refinement an
increase of the volume of the Universe is due to the formation of
the new lattice vertices, while the spin labels of the spin
network remain constant. It was shown in Ref. \cite{MS} that this
choice is the only lattice refinement model with a non ambiguous
and correct classical limit.

Our model applies the $\bar{\mu}$-scheme (see above) and
$\text{Ar}_{\Box_{ij}}= \lambda^2 = $ const, where $\lambda$ is a
free parameter. Because of this choice the nonstandard and the
standard LQC give the same predictions at the semiclassical
level.

\acknowledgments

We thank  Stanis{\l}aw G{\l}azek and Lech Szymanowski for helpful discussions.
JM has been supported by the Polish Ministry of Science and Higher Education
grant N N203 386437.

\end{document}